\documentclass[tighten,twocolumn]{aastex63}

\usepackage{graphicx}
\usepackage{ulem}
\usepackage{amsmath}
\usepackage{wrapfig}
\usepackage[thinlines]{easytable}
\usepackage{tikz}
\usepackage{hyperref}

\newcolumntype{P}[1]{>{\centering\arraybackslash}p{#1}}

\begin{document}
\shorttitle{neutron star tidal disruption events}
\shortauthors{Kremer et al.}

\title{Formation of Low-mass Black Holes and Single Millisecond Pulsars in Globular Clusters}

\correspondingauthor{Kyle Kremer}
\email{kkremer@caltech.edu}

\author[0000-0002-4086-3180]{Kyle Kremer}
\altaffiliation{NSF Astronomy \& Astrophysics Postdoctoral Fellow}
\affiliation{TAPIR, California Institute of Technology, Pasadena, CA 91125, USA}
\affiliation{The Observatories of the Carnegie Institution for Science, Pasadena, CA 91101, USA}

\author[0000-0001-9582-881X]{Claire S.\ Ye}
\affiliation{Center for Interdisciplinary Exploration \& Research in Astrophysics (CIERA) and Department of Physics \& Astronomy \\ Northwestern University, Evanston, IL 60208, USA}

\author[0000-0003-4412-2176]{Fulya K{\i}ro\u{g}lu}
\affiliation{Center for Interdisciplinary Exploration \& Research in Astrophysics (CIERA) and Department of Physics \& Astronomy \\ Northwestern University, Evanston, IL 60208, USA}

\author[0000-0002-7444-7599]{James C.\ Lombardi Jr.}
\affiliation{Department of Physics, Allegheny College, Meadville, Pennsylvania 16335, USA}

\author[0000-0001-5799-9714]{Scott M.\ Ransom}
\affiliation{NRAO, 520 Edgemont Road, Charlottesville, VA 22903, USA}

\author[0000-0002-7132-418X]{Frederic A.\ Rasio}
\affiliation{Center for Interdisciplinary Exploration \& Research in Astrophysics (CIERA) and Department of Physics \& Astronomy \\ Northwestern University, Evanston, IL 60208, USA}

\begin{abstract}
Close encounters between neutron stars and main-sequence stars occur in globular clusters and may lead to various outcomes. Here we study encounters resulting in tidal disruption of the star. Using $N$-body models, we predict the typical stellar masses in these disruptions and the dependence of the event rate on host cluster properties. We find that tidal disruption events occur most frequently in core-collapsed globular clusters and that roughly $25\%$ of the disrupted stars are merger products (i.e., blue straggler stars). Using hydrodynamic simulations, we model the tidal disruptions themselves (over timescales of days) to determine the mass bound to the neutron star and the properties of the accretion disks formed. In general, we find roughly $80-90\%$ of the initial stellar mass becomes bound to the neutron star following disruption.
Additionally, we find that neutron stars receive impulsive kicks of up to about $20\,$km/s as a result of the asymmetry of unbound ejecta; these kicks place these neutron stars on elongated orbits within their host cluster, with apocenter distances well outside the cluster core. Finally, we model the evolution of the (hypercritical) accretion disks on longer timescales (days to years after disruption) to estimate the accretion rate onto the neutron stars and accompanying spin-up. As long as $\gtrsim1\%$ of the bound mass accretes onto the neutron star, millisecond spin periods can be attained. We argue the growing numbers of {\em isolated\/} millisecond pulsars observed in globular clusters may have formed, at least in part, through this mechanism. In the case of significant mass growth, some of these neutron stars may collapse to form low-mass ($\lesssim3\,M_{\odot}$) black holes. 

\vspace{1cm}
\end{abstract}

\section{Introduction}

Close stellar encounters have long been understood to be a prominent feature of dense star clusters. Encounters involving neutron stars have drawn particular interest since the 1980s when the first globular cluster millisecond pulsars (MSPs) were discovered \citep{Lyne1987}. In the standard picture, cluster MSPs are thought to form in low-mass X-ray binaries where the neutron star is spun up through accretion of material from its companion \citep[e.g.,][]{PhinneyKulkarni1994}. These binaries are expected to be assembled primarily through dynamical encounters where a neutron star is dynamically exchanged into a binary \citep[e.g.,][]{Sigurdsson1995,CamiloRasio2005,Ye2019}. Once formed, the neutron star binary is hardened by subsequent dynamical encounters until the star fills its Roche lobe and begins transferring mass onto the neutron star and spinning it up.

There are now more than 200 MSPs observed in Milky Way clusters \citep{psr_catalog}, many of which are found in binaries, as expected for the standard low-mass X-ray binary formation scenario. However, in recent years, a growing number of \textit{isolated} MSPs have been observed in globular clusters. The ratio of isolated to binary MSPs is particularly pronounced in core-collapsed globular clusters. 
In all core-collapsed clusters with at least one known MSP, \textit{roughly 80\% of observed MSPs are found without binary companions} \citep[on average;][]{psr_catalog}.
This result defies the standard dynamical exchange formation scenario for MSPs which naively suggests the opposite trend: the densest clusters should in fact feature an increased number of binary MSPs \citep[e.g.,][]{VerbuntHut1987,Pooley2003}. In light of this, alternative formation scenarios for MSPs may be necessary.

In old core-collapsed clusters, the inner regions are dominated by massive main-sequence stars, 
white dwarfs, and neutron stars \citep[e.g.,][]{Kremer2021b}. The high densities within the centers of core-collapsed clusters lead naturally to an increased rate of close stellar encounters of these objects, and therefore, an increased rate of stellar collisions and tidal disruptions \citep[e.g.,][]{HeggieHut2003}. A number of studies have shown that close tidal interactions involving specifically neutron stars and stars may lead to the formation of compact binaries \citep[e.g.,][]{Fabian1975,Ray1987,Ivanova2005,Ye2022}. Even closer encounters inevitably lead to collisions of neutron stars and stars \citep[e.g.,][]{KrolikMeiksinJoss1984,RasioShapiro1991,Lombardi2006,Perets_2016,Kremer2019c}. Previous analyses have shown that such collisions may lead to common envelope-like events that may ultimately result in Thorne-$\dot{\rm Z}$ytkow objects \citep{ThorneZytkow1977} or, if significant accretion and spin up occurs, MSPs \citep[e.g.,][]{Davies1992,DaviesBenz1995,Lee1996,CamiloRasio2005}.

Here we explore the possibility of MSP formation through accretion onto neutron stars following the tidal disruption of main-sequence stars. Our analysis proceeds in three steps: In Section \ref{sec:CMC}, we analyze a suite of realistic $N$-body models that span the full parameter space of the Milky Way globular clusters to predict the rates and demographics of these tidal disruption events (TDEs). Using models tuned to NGC~6752, NGC~6624, and 47~Tuc, we predict the number of TDEs in clusters with known isolated MSP populations. Motivated by the $N$-body results, in Section \ref{sec:SPH} we perform hydrodynamic simulations of a few representative encounters to investigate the outcome of the TDEs. Finally, motivated by the hydrodynamics, in Section \ref{sec:accretion}, we present a simple analytic model for the long-term evolution of the accretion disks formed. Incorporating key uncertainties, we predict the total mass and angular momentum accreted by the neutron star and address the key question of whether millisecond spin periods are attainable. We discuss our results and conclude in Section \ref{sec:discussion}.

\section{Rates and Demographics of TDEs from N-body models}
\label{sec:CMC}

\begin{figure}
    \centering
    \includegraphics[width=\columnwidth]{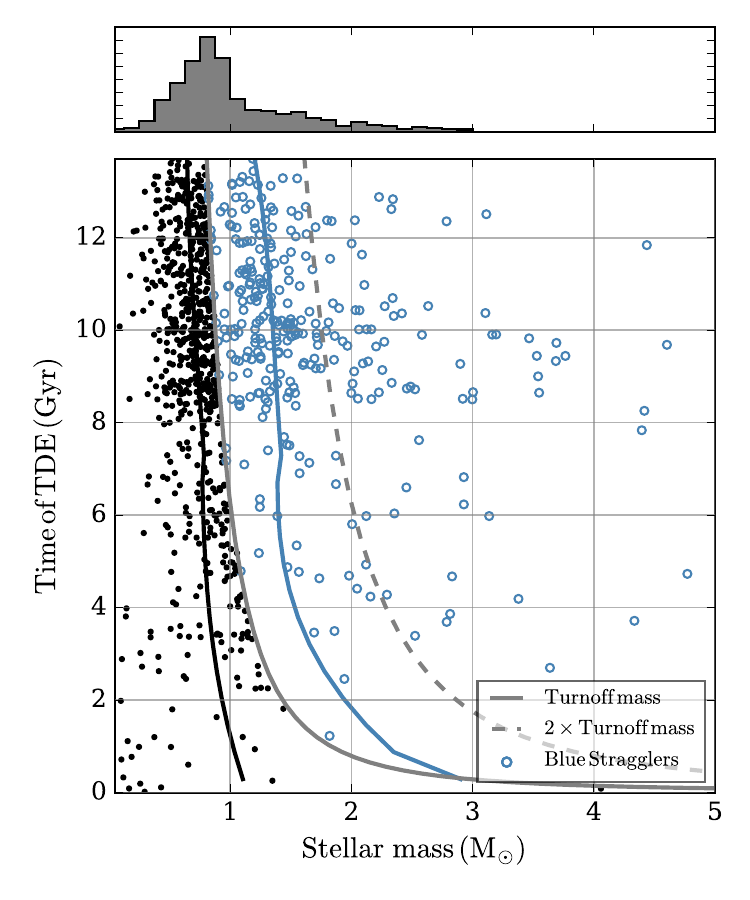}
    \caption{\footnotesize Time of TDE versus stellar mass for all main-sequence star+neutron star TDEs occurring in our \texttt{CMC} cluster models. Black circles indicate standard main-sequence stars, open blue circles indicate stars with masses above the turnoff mass (shown as a solid gray curve) which are likely observationally identified as blue straggler stars. The solid black (blue) curves show the median disrupted mass versus time for the main-sequence star (blue straggler star) TDEs.}
    \label{fig:mass}
\end{figure}

\begin{deluxetable*}{l|l|c|c|c|c|c|c|c|c||c}
\tabletypesize{\footnotesize}
\tablewidth{0pt}
\tablecaption{Neutron star+main-sequence star TDE counts from \texttt{CMC} cluster models \label{table:CMC}}
\tablehead{
    \colhead{} &
	\colhead{$^1$Cluster model} &
	\colhead{$^2 r_{v,i}$} &
	\colhead{$^3 M_{\rm cl}$} &
	\colhead{$^4 r_c$} &
	\colhead{$^5 r_h$} &
	\colhead{$^6 n$} &
	\colhead{$^7$CC?} &
	\colhead{$^{8}$\# NS+MS TDEs} &
	\colhead{$^{9}$\# NS+BSS TDEs} &
	\colhead{$^{10}$\# isolated MSPs}\\
	\colhead{} &
	\colhead{} &
	\colhead{pc} &
	\colhead{$\times10^5\,M_{\odot}$} &
	\multicolumn{2}{c}{pc} &
	\colhead{$\rm{pc}^{-3}$} &
	\colhead{} &
	\colhead{} &
	\colhead{} &
	\colhead{[observed]}
}
\startdata
1 & \textsc{n8-rv0.5-rg8-z0.1} & 0.5 & 1.99 & 0.17 & 2.00 & $2.2\times10^5$ & Y & 59 & 30 & 58 \\
2 & \textsc{n8-rv1-rg8-z0.1} & 1 & 2.21 & 0.61 & 2.18 & $1.1\times10^4$ & N & 2 & 0 & 1 \\
3 & \textsc{n8-rv2-rg8-z0.1} & 2 & 2.31 & 2.81 & 4.06 & $700$ & N & 0 & 0 & 0 \\
4 & \textsc{n8-rv4-rg8-z0.1} & 4 & 2.34 & 4.66 & 7.26 & $140$ & N & 0 & 0 & 0 \\
\hline
5 & ``NGC6752'' & 0.5 & 1.85 & 0.14 & 2.53 & $3\times10^5$ & Y & 125 & 37 & 121 [8] \\
6 & ``NGC6624'' & 1 & 1.79 & 0.18 & 1.08 & $9.7\times10^4$ & Y & 22 & 12 & 22 [7] \\
7 & ``47~Tuc'' & 4 & 9.68 & 0.8 & 7.0 & $7.3\times10^4$ & N & 19 & 3 & 19 [10] \\
\hline
 & Total (``Milky Way'')& - & - & - & - & - & - & 988 (1500) & 304 (500) & 950 (1400) [98] \\
\enddata
\tablecomments{\footnotesize Total number of TDEs identified in various models from the \texttt{CMC Cluster Catalog}. In column 2 we list the initial virial radius for each model \citep{Kremer2019a,Ye2022}. In columns 3-6 we list the present-day cluster mass, core radius, half-light radius, and central density (within core radius) of each model. In column 7, we denote whether the cluster model is core-collapsed at present. In column 10, we list the total number of TDEs in each model that lead to sufficient spin-up to form isolated MSPs (assuming $s=0.2$ as discussed in Section \ref{sec:accretion}) and, in brackets, we list the observed number of isolated MSPs, where relevant. In the rows 1-4, we list TDE counts for four ``typical'' cluster models with varying virial radius, $r_v$. In rows 5-7, we show models that match well three specific clusters. In the last row, we list the total number of TDEs identified in the full cluster catalog and the inferred total number of TDEs expected in the full Milky Way globular cluster system (shown in parentheses).}
\end{deluxetable*}

To compute the numbers of TDEs in typical clusters, we use the $N$-body models from our \texttt{CMC Cluster Catalog} \citep{Kremer2020} which were computed using \texttt{CMC} \citep{Rodriguez2022}, a H\'{e}non-type Monte Carlo code that includes prescriptions for various physical processes relevant to dense star clusters including two-body relaxation, stellar and binary evolution \citep[computed using \texttt{COSMIC};][]{Breivik2020}, and direct integration of small-$N$ resonant encounters \citep{Fregeau2007,Rodriguez2018}. A number of parameters are varied within the \texttt{CMC Catalog} namely the initial cluster mass, initial virial radius, metallicity\footnote{As we are specifically interested here in only those clusters that are sufficiently old to have undergone core collapse, we exclude the solar metallicity models published as part of the \texttt{CMC Cluster Catalog} and only examine TDEs occurring in models with $Z=0.01Z_{\odot}$ and $0.1Z_{\odot}$, most typical of the Milky Way globular clusters.}, and radial position within the Galactic potential. Altogether, this catalog effectively spans the full parameter space of the Milky Way globular clusters and captures the formation of a variety of astrophysical objects such as gravitational-wave, X-ray binaries, (millisecond) pulsars, cataclysmic variables, and blue straggler stars.

In \texttt{CMC}, we record a TDE whenever a neutron star passes within the tidal disruption radius of a nearby main-sequence star: $r_T=(M_{\rm{NS}}/M_{\star})^{1/3}R_{\star}$, where $M_{\rm NS}$ is the neutron star mass and $M_{\star}$ and $R_{\star}$ are the mass and radius of the star. In the case of $M_{\star}>M_{\rm NS}$, $r_T<R_{\star}$ and the relevant minimum pericenter distance for disruption is simply the stellar radius.

In Figure \ref{fig:mass} we show the stellar mass and disruption time for all TDEs identified in our suite of models. Black scatter points denote ``normal'' main-sequence stars with mass below the turn-off mass (indicated by the solid gray curve). Roughly $75\%$ of identified TDEs fall into this category. Open blue circles denote stars with mass above the turn-off mass. These stars (which constitute roughly $25\%$ of all identified TDEs) were formed from previous stellar collisions/mergers and would observationally be identified as blue straggler stars \citep[BSSs; e.g.,][]{Sandage1953}. We also show as a dashed gray curve the boundary marking twice the turn-off mass. Stars to the right of this dashed curve were formed through two or more stellar collisions. As shown the disrupted stellar masses range from roughly $0.1\,M_{\odot}$ (the assumed lower limit of the stellar mass function) to roughly $5\,M_{\odot}$. At very early times, there are also a handful of TDEs with $M_{\star}>5\,M_{\odot}$, which are not plotted here since they are rare. The median mass of all disrupted stars is roughly $0.9\,M_{\odot}$ (averaged over all time). For BSS (main-sequence star) TDEs, the median disrupted mass is roughly $1.4\,M_{\odot}$ ($0.7\,M_{\odot}$). In Figure~\ref{fig:mass}, we show as solid blue (black) curves the median mass of BSS (main-sequence star) TDEs versus time. Finally, the majority of TDEs ($\gtrsim80\%$) occur at late times ($t>8\,$Gyr), after their host clusters' stellar-mass black hole populations have been mostly depleted \citep[for discussion, see][]{Kremer2020}.

\begin{figure*}
    \centering
    \includegraphics[width=0.85\linewidth]{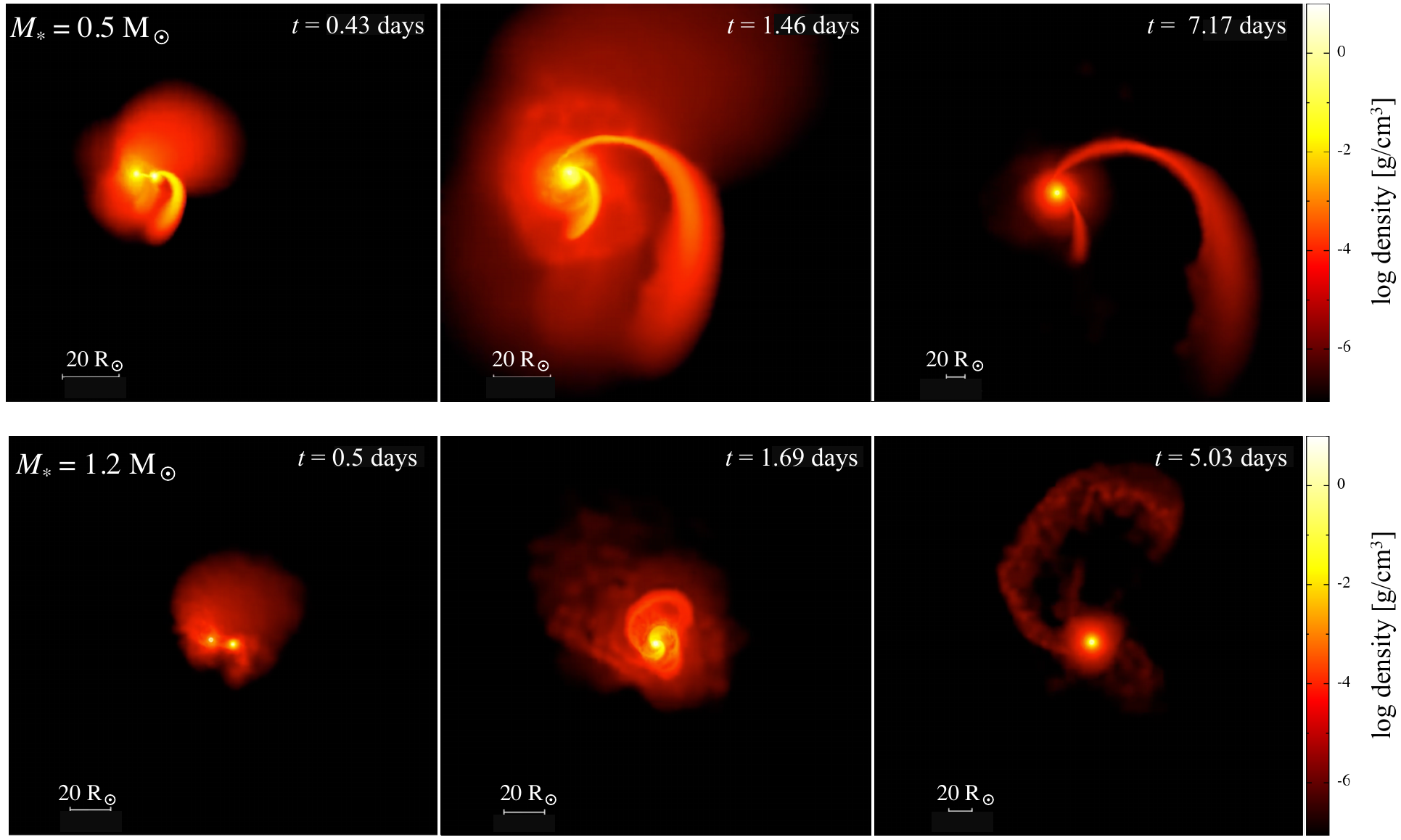}
    \caption{\footnotesize Hydrodynamic evolution of two fiducial SPH simulations. On top, we show a $M_{\star}=0.5\,M_{\odot}$ (modeled as an $n=1.5$ polytrope; simulation 3 in Table \ref{table:sims}) and on bottom, we show a $M_{\star}=1.2\,M_{\odot}$ (a ``blue straggler star'' modeled as an Eddington standard model; simulation 6 in Table \ref{table:sims}). In both simulations, $r_p=r_T$ is adopted. In the $M_{\star}=0.5\,M_{\odot}$ ($M_{\star}=1.2\,M_{\odot}$) case, roughly $0.45\,M_{\odot}$ ($1.1\,M_{\odot}$) is bound to the neutron star at the end of the simulation. Links to animations for these simulations (and others) are included in Table \ref{table:sims}. The animation corresponding to the top (bottom) panel covers the simulation from $t = -0.3-7.5\,$d ($t = -0.5-6.3\,$d). In these two animations, the star is partially disrupted during the first pericenter passage. The partially stripped star becomes bound to the neutron star and returns for subsequent passages before ultimately being destroyed completely.
    }
    \label{fig:SPH1}
\end{figure*}

In Table \ref{table:CMC} we list the total number of TDEs as well as BSS TDEs in four representative models from the \texttt{CMC Cluster Catalog} with initial stellar number of $N=8\times10^5$, Galactocentric distance of $8\,$kpc, metallicity of $Z=0.1Z_{\odot}$ and four different initial virial radii from $0.5-4\,$pc.\footnote{For a list of neutron star TDEs (and TDEs/collisions involving other stellar types) in the complete set of models in the \texttt{CMC Cluster Catalog}, see the Appendix of \citet{Kremer2020}.} As described in \citet{Kremer2019a}, the initial virial radius (which sets the initial density of the model) determines whether or not the cluster undergoes core collapse by the present-day age ($t\sim12\,$Gyr). For $N=8\times 10^5$, only models with $r_v=0.5\,$pc have reached core collapse by this time. As a result, the $r_v=0.5\,$pc model yields by far the most TDEs of the four.

In addition to the four representative models, we also list the \texttt{CMC Catalog} models that most effectively match the surface brightness and velocity dispersion profiles \citep[following the method of][]{Rui2021} of two core-collapsed clusters with large numbers of known isolated MSPs: NGC~6752 and NGC~6624 \citep{psr_catalog}. We also provide this same information for our \texttt{CMC} model for 47~Tuc \citep{Ye2022}. Although 47~Tuc is not core collapsed, it is sufficiently massive and dense to still yield a large number of TDEs (and observed isolated MSPs). In the last row of Table \ref{table:CMC}, we list the total number of TDEs in the full set of models, and the inferred total (in parentheses) for the full Milky Way cluster population determined by scaling up the set of models to match the total Milky Way cluster mass.

In typical core-collapsed clusters like NGC~6752, we predict an event rate of neutron star+star TDEs of up to roughly $10\,\rm{Gyr}^{-1}$, averaged over the full ($\sim12\,$Gyr) lifetime. At late times ($t>8\,$Gyr) after core collapse has occurred, we find event rates of up to roughly $30\,\rm{Gyr}^{-1}$ per cluster. For massive non-core-collapsed clusters like 47~Tuc (and Terzan~5), we predict a TDE rate of roughly $2\,\rm{Gyr}^{-1}$ over the lifetime of the cluster. For lower-mass non-core-collapsed clusters (e.g., simulations \texttt{2-4} in Table \ref{table:CMC}), we predict an event rate of at most $0.1\,\rm{Gyr}^{-1}$. For our inferred Milky Way population \citep[in which roughly $20\%$ of globular clusters reach core collapse by present day;][]{Harris1996}, we estimate a total TDE event rate of roughly $100\,\rm{Gyr}^{-1}$.

These rates are comparable to those found in previous studies. For instance, \citet{Sigurdsson1995} showed neutron star+star collisions occur most frequently in the densest clusters ($n>10^5\,\rm{pc}^{-3}$; comparable to that expected for core-collapsed clusters). In such clusters, this study estimated rates of roughly 100 per cluster lifetime, similar to our findings. Furthermore, this study predicted that roughly 75\% of MSPs in these dense clusters should be single. Similarly, \citet{DaviesHansen1998} estimated neutron star collision rates of roughly $100\,\rm{Gyr}^{-1}$ for clusters with $n>10^5\,\rm{pc}^{-3}$, corresponding to a rate of roughly $1000\,\rm{Gyr}^{-1}$ in the full Milky Way cluster population, consistent with our estimate given that they did not take into account the fact that core-collapsed clusters likely only spend a fraction of their lives in a core-collapsed state with extreme central densities. Finally, \citet{Hansen1998} used the inferred MSP birth rate in clusters to estimate a neutron star+star collision rate of roughly $10-10^4\,\rm{Gyr}^{-1}$ in the Milky Way, also consistent with our estimate.

\section{Hydrodynamic Evolution}
\label{sec:SPH}

Motivated by the overall TDE demographics from our $N$-body models, we now explore the hydrodynamic evolution of these TDEs for a few representative cases. This work extends upon previous studies on this topic \citep[e.g.,][]{Davies1992,Rasio1993,Ivanova2005,Lombardi2006}. In order to explore the hydrodynamic outcomes, we use the smoothed particle hydrodynamics (SPH) code \texttt{StarSmasher} \citep{RasioThesis,Evghenii_2018}. To model close encounters of neutron stars and main-sequence stars, we follow the method described in \citet{Kremer2022} where the neutron star is treated as a ``point particle'' and interacts with the 200k SPH particles of the star only via softened gravity.
We perform eight SPH simulations, which are summarized in Table \ref{table:sims}. In all cases, we adopt $v_{\infty}=0\,$km/s, representative of nearly parabolic encounters expected in typical globular clusters. Additionally, we assume a fixed neutron star mass of $1.2\,M_{\odot}$, typical for neutron stars formed through electron capture supernovae, expected to be the most common formation channel for neutron stars retained in globular clusters \citep[e.g.,][]{Ye2019}.

For the first three simulations, we model the encounter of a $0.5\,M_{\odot}$ M-dwarf modeled as an $n=1.5$ polytrope, governed by a polytropic equation of state with adiabatic index $\Gamma=5/3$, interacting with a neutron star at three pericenter distances: $r_p=[0.5,0.75,1]\times r_T$. These $M_{\star}=0.5\,M_{\odot}$ cases are representative of the low mass TDEs expected to occur (see Figure \ref{fig:mass}). Roughly 40\% of the TDEs in our \texttt{CMC} models feature a stellar mass less than $0.8\,M_{\odot}$ (i.e., M- and K-dwarfs). For the second set of three simulations, we model the case of a $1.2\,M_{\odot}$ main-sequence star modeled as an Eddington standard stellar model \citep[$n=3$ polytrope index with an equation of state incorporating both ideal gas and radiation pressure, as in][]{Rasio1993_Gamma}, representative of the BSS TDEs shown in Figure \ref{fig:mass}. We also include two additional simulations (simulations 7 and 8 in the table) adopting $M_{\star}=0.8\,M_{\odot}$ and $M_{\star}=2\,M_{\odot}$. The $M_{\star}=0.8\,M_{\odot}$ simulation explores the case of the disruption of a near turn-off mass main-sequence star representative of the most common disrupted star (see Section~\ref{sec:CMC}). The $M_{\star}=2\,M_{\odot}$ simulation explores the rarer case of even more massive stellar disruptions (less than 8\% of the TDEs identified in Section \ref{sec:CMC} have stellar masses of $2\,M_{\odot}$ or more). For each of these two extra simulations, we adopt $r_p=r_T$ and the Eddington standard stellar model. Finally, the initial radii of the $0.5\,M_{\odot}$, $0.8\,M_{\odot}$, $1.2\,M_{\odot}$, and $2\,M_{\odot}$ stellar models are $0.6\,R_{\odot}$, $0.8\,R_{\odot}$, $1.2\,R_{\odot}$, and $1.5\,R_{\odot}$, respectively.

The three pericenter distances chosen span a reasonable range of encounter types: for $r_p/r_T=0.5$, the neutron star penetrates well into the stellar radius, while the $r_p/r_T=1$ case is a more classic tidal disruption. We do not simulate even more distant encounters where little mass is stripped and stable neutron star+main-sequence star binaries may form through tidal capture \citep[e.g.,][]{Fabian1975,Teukolsky_1977}, although these more distant encounters may well play an important role in the formation of binary MSPs \citep{Ye2022}. We discuss this further in Section \ref{sec:discussion}.

\begin{deluxetable*}{l|c|c|c|cc|cccc|c|c|c|c}
\tabletypesize{\footnotesize}
\tablewidth{0pt}
\tablecaption{List of SPH simulations performed \label{table:sims}}
\tablehead{
	\colhead{} &
	\colhead{$^1M_{\star}$} &
	\colhead{$^2R_{\star}$} &
	\colhead{$^3r_p/r_T$} &
	\colhead{$^4M_{\rm{bound,NS}}$} &
	\colhead{$^5M_{\rm{ej}}$} &
	\colhead{$^6J_{\rm{disk}}$} &
	\colhead{$^{7}R_{\rm{disk}}$} &
	\colhead{$^{8}\Omega_{\rm{disk}}$} &
	\colhead{$^{9}t_v$} &
	\colhead{$^{10}\dot{M}$} &
	\colhead{$^{11}\Delta t$} &   
	\colhead{$^{12}v_{\rm{NS}}$} &
	\colhead{}\\
	\colhead{} &
	\colhead{$M_{\odot}$} &
	\colhead{$R_{\odot}$} &
	\colhead{} &
	\multicolumn{2}{c}{$M_{\odot}$} &
	\colhead{$M_{\odot} R_{\odot}^2 \rm{d}^{-1}$} &
	\colhead{$R_{\odot}$} &
	\colhead{$\rm{d}^{-1}$} &
	\colhead{$\rm{d}$} &
	\colhead{$M_{\odot}\,\rm{yr}^{-1}$} &
	\colhead{$\rm{d}$}&
	\colhead{km/s}&
	\colhead{}
}
\startdata
1 & 0.5 & 0.6 & 0.5 & 0.439 & 0.061 & 20.89 & 4.55 & 6.39 & 1.57 & 102.4 & N/A & 15.98 & \href{https://sites.northwestern.edu/kremerastronomy/files/2022/05/NS_r0.5q2.4.mov}{video}\\
2 & 0.5 & 0.6 & 0.75 & 0.389 & 0.111 & 22.75 & 8.93 & 2.04 & 4.91 & 28.9 & 0.31 & 6.12 & \href{https://sites.northwestern.edu/kremerastronomy/files/2022/05/NS_r0.75q2.4.mov}{video}\\
3 & 0.5 & 0.6 & 1.0 & 0.447 & 0.053 & 27.89 & 8.97 & 2.15 & 4.64 & 35.1 & 0.65 & 2.46 & \href{https://sites.northwestern.edu/kremerastronomy/files/2022/05/NS_r1.00q2.4.mov}{video}\\
\hline
4 & 1.2 & 1.5 & 0.5 & 0.937 & 0.263 & 35.02 & 2.93 & 4.35 & 2.3 & 148.7 & 0.15 & 23.52 & \href{https://sites.northwestern.edu/kremerastronomy/files/2022/03/r0.50mstar1.2mns1.2_200K-1.mov}{video} \\
5 & 1.2 & 1.5 & 0.75 & 1.056 & 0.144 & 56.41 & 2.82 & 6.73 & 1.49 & 259.2 & 0.45 & 9.69 & \href{https://sites.northwestern.edu/kremerastronomy/files/2022/03/r0.75mstar1.2mns1.2_200K-1.mov}{video} \\
6 & 1.2 & 1.5 & 1.0 & 1.107 & 0.093 & 68.32 & 3.56 & 4.87 & 2.05 & 196.9 & 1.52 & 3.72 & \href{https://sites.northwestern.edu/kremerastronomy/files/2022/02/NS_r1.00mstar1.2mns1.2_200K.mov}{video}\\
\hline
7 & 0.8 & 0.8 & 1.0 & 0.719 & 0.081 & 43.09 & 3.27 & 5.62 & 1.78 & 147.5 & 0.79 & 1.89 & \href{https://sites.northwestern.edu/kremerastronomy/files/2022/06/NS_r1.00mstar0.8mns1.2.mov}{video}\\
\hline
8 & 2.0 & 1.2 & 1.0 & 1.735 & 0.265 & 82.39 & 3.46 & 3.97 & 2.52 & 251.7 & 1.1 & 11.56 & \href{https://sites.northwestern.edu/kremerastronomy/files/2022/03/r1.00mstar2mns1.2_200K.mov}{video}
\enddata
\tablecomments{\footnotesize Outcomes of all SPH simulations. Columns 1 and 2 show initial stellar mass and radius. Column 3 shows pericenter distance of encounter in units of tidal disruption radius. We report all properties in columns 4-9 after the final pericenter passage. $\dot{M}=M_{\rm disk}/t_v$ (column 10) shows the characteristic peak mass inflow rate. $\Delta t$ (column 11) denotes the time elapsed between the initial pericenter passage and the final pericenter passage (when the star is fully disrupted). In column 12 we list the final velocity (reported at the end of the simulation) of the neutron star (attained from the impulsive kick imparted by the asymmetric ejecta mass).
}
\end{deluxetable*}

In nearly all simulations performed, the star is partially disrupted during the first pericenter passage, becomes bound to the neutron star (i.e., is tidally captured), and is ultimately disrupted fully after one or more additional passages. The only exception is simulation~1, where the star is disrupted fully on the first pericenter passage due to the relatively high penetration depth of the encounter and relatively high mass ratio. As expected, the time between the initial pericenter passage and full disruption of the star (shown as $\Delta t$ in column~11 of Table~\ref{table:sims}) increases with increasing $r_p/r_T$.

In columns~4 and~5 of Table~\ref{table:sims}, we list the total mass bound to the neutron star and total mass ejected from the system after final disruption. In all cases, we find that roughly $80-90\%$ of the initial stellar mass becomes bound to the neutron star, while the remaining roughly $10-20\%$ of mass becomes unbound from the system.
In columns~6, 7, and~8 we report, respectively, the total angular momentum of the material bound to the neutron star, characteristic disk radius, and characteristic disk angular frequency. These quantities are computed as in Equations~(15), (16), and~(17) of \citet{Kremer2022}. In column 9, we report the characteristic viscous accretion timescale $t_v\approx (\alpha h^2 \Omega_{\rm disk})^{-1}$ of the material bound to the neutron star (where $h\approx1$ is the disk scale height ratio and $\alpha \approx 0.1$ is the assumed disk viscosity). All quantities in columns~4--9 are reported at a time $2 t_{\rm orb}$ (where $t_{\rm orb}$ is the orbital period of material bound to the neutron star) after the final pericenter passage (when the star is completely disrupted). As discussed in \citet{Kremer2022}, $2t_{\rm orb}$ is chosen to ensure a disk has had sufficient time to form. On longer timescales ($t\gtrsim$days), the hydrodynamic evolution is governed by the accretion process of the disk, which is not modeled in our SPH set up. Importantly however, the viscous accretion times estimated here are longer than the typical time elapsed between the first pericenter passage and the final disruption of the star (column 11 of Table \ref{table:sims}). This indicates that significant accretion onto the neutron star is unlikely to occur on the timescales of our SPH simulations ($t\lesssim$days), justifying our assumption to ignore accretion in the SPH modelling. We discuss the possible long-term outcome of the evolving disks in Section~\ref{sec:accretion}.

In column 10 of Table \ref{table:sims}, we show the characteristic peak mass inflow rate of the disks, defined as $\dot{M}=M_{\rm disk}/t_v$ where $M_{\rm disk}$ and $t_v$ are obtained directly from the SPH simulations (column 4 and 9, respectively). We discuss the accretion process further in Section~\ref{sec:accretion}. Finally, in column 12 of the table we report the final velocity of the neutron star. As described in \citet{Kremer2022} in the context of black hole TDEs, the compact object is expected to receive a dynamical ``kick'' as a result of the impulse from material ejected to infinity. For more penetrating encounters, the geometry of the unbound ejecta becomes increasingly asymmetric and as a result, the velocity of the kick increases for encounters that are more nearly head-on. For the TDEs modeled here, we find kick velocities ranging from roughly $2-20\,$km/s. While unlikely to be sufficient to eject the neutron stars from their host cluster, these velocities are likely sufficient to kick the neutron stars onto elongated orbits within their host. We return to this point in Section \ref{sec:discussion}.

To illustrate the key effects, we show in Figure~\ref{fig:SPH1} the hydrodynamic evolution of models~3 and~6. We also provides links to videos of all simulations in the final column of Table~\ref{table:sims}.\footnote{Figure \ref{fig:SPH1} and accompanying animations were created using the \texttt{SPLASH} visualization software \citep{Splash2007}.}

\section{Accretion, spin up, and formation of millisecond pulsars}
\label{sec:accretion}

\begin{figure*}
    \centering
    \gridline{\fig{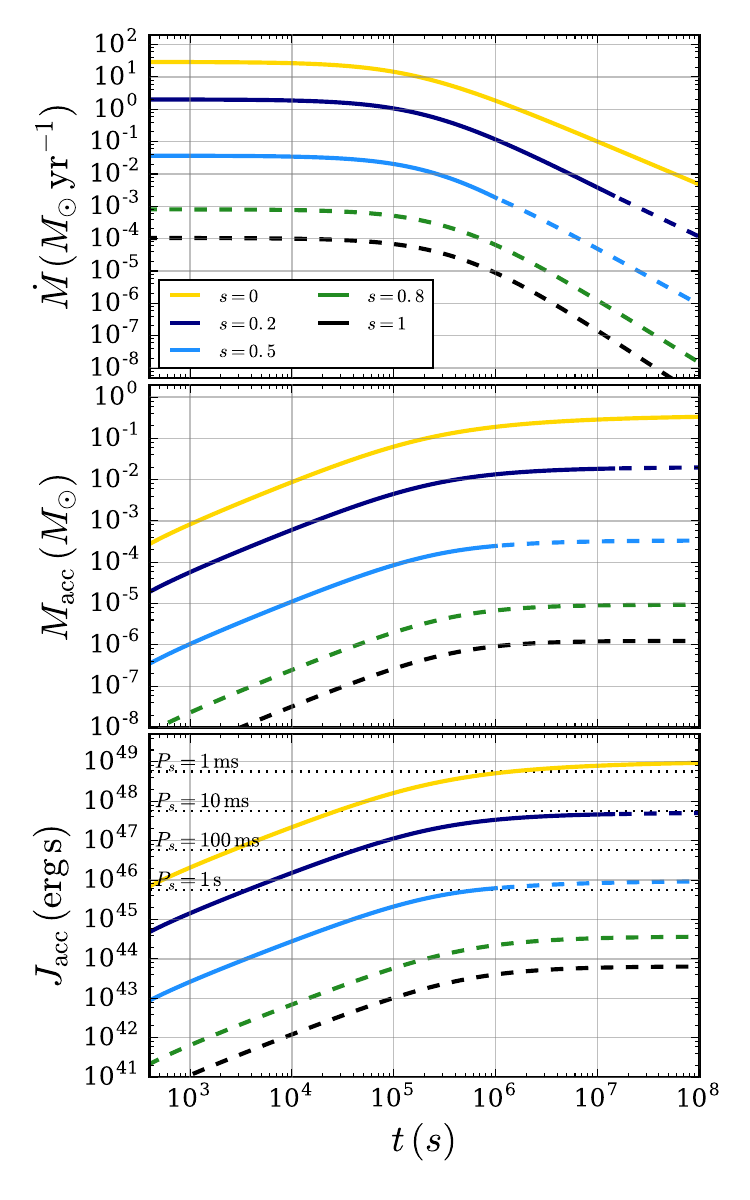}{0.45\textwidth}{}
          \fig{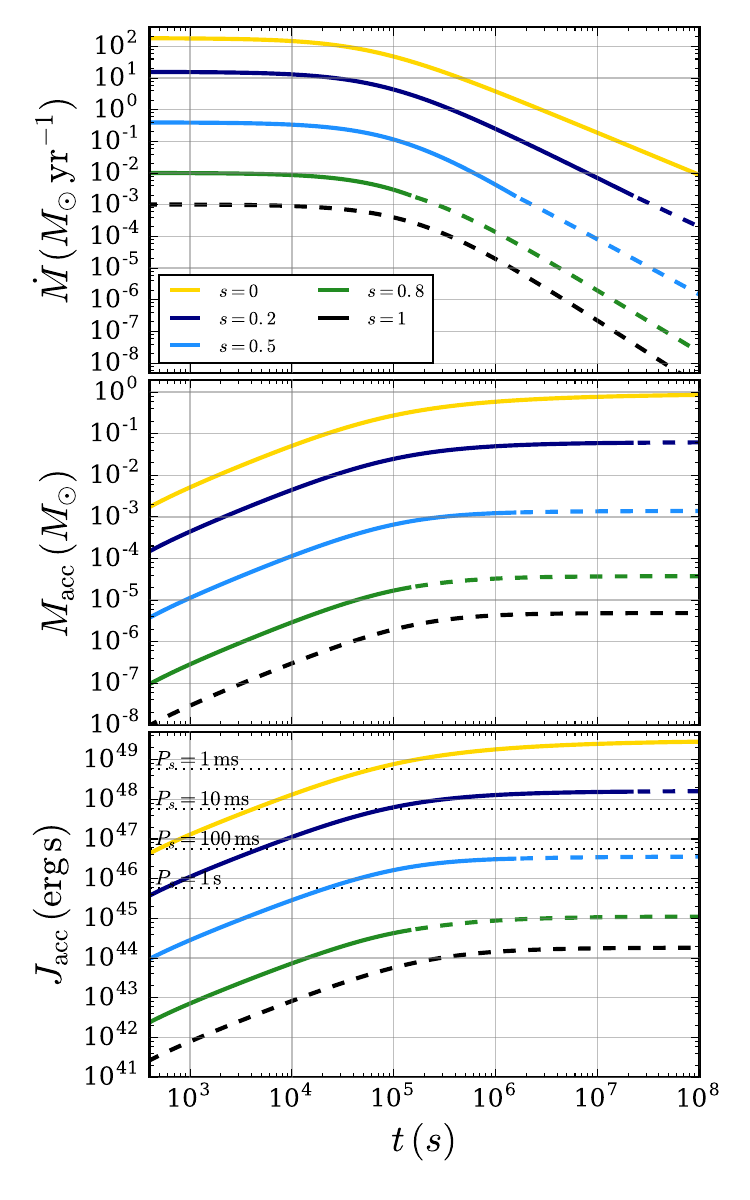}{0.45\textwidth}{}}
    \caption{\footnotesize Evolution of accretion disks for first $\sim\,$year after tidal disruption. In the left (right) column we show the evolution of a $0.4\,M_{\odot}$ ($1\,M_{\odot}$) disk. Top panels show the accretion rate onto the neutron star, middle panels show the total (cumulative) mass accreted, and bottom panels show the total spin angular momentum accreted. Solid curves indicate where the mass transfer rate is sufficiently high to ``smother'' the magnetic field (in this case the field -- assumed here to be $10^{11}$\,G -- does not play an important role in the accretion flow) while dashed lines indicate the point in evolution where magnetic energy density dominates the accretion flow. Finally, as horizontal dotted lines in the bottom panels, we show neutron star spin angular momentum values for a few representative spin periods.}
    \label{fig:disk}
\end{figure*}

Motivated by the disk features (e.g., mass, radius, and angular momentum) predicted from the SPH simulations, we now build a simple model for the disk evolution and discuss possible accretion rates onto the neutron star. For the material bound to the neutron star, we find typical masses of roughly $0.5-1\,M_{\odot}$ and viscous accretion times of roughly $1-5\,$d (columns 4 and 9 of Table \ref{table:sims}). This implies a characteristic mass inflow rate of roughly $30-300\,M_{\odot}\,\rm{yr}^{-1}$ (column 10 of Table~\ref{table:sims}), orders of magnitude above the standard Eddington accretion limit of roughly $10^{-8}\,M_{\odot}\,\rm{yr}^{-1}$ for typical neutron star assumptions \citep[e.g.,][]{PhinneyKulkarni1994}.
The possibility of so-called ``hypercritical'' accretion has been considered in a variety of contexts including accretion during supernovae \citep[e.g.,][]{Colgate1971,Zeldovich1972}, as a feature of gamma-ray burst models \citep[e.g.,][]{Popham1999, Lee2005}, and for accretion in a common envelope \citep[e.g.,][]{Chevalier1993,Brown1995,Terman1995,FryerBenzHerant1996,BetheBrown1998,Hansen1998,MacLeod2015}. The key is there must exist a way for the accretion energy to be radiated away without impeding the flow and limiting the mass accretion rate onto the neutron star. In principle, this may occur in several ways. In the case of disk-like accretion flows where the inward mass flow predominately occurs in the plane of the disk (similar to what is expected for off-axis tidal disruptions like those modeled with our SPH simulations described in the previous section), the accretion energy may flow relatively freely out low-density polar regions \citep[e.g.,][]{Frank2002}. Even for relatively spherical Bondi-Hoyle type \citep{BondiHoyle1944} accretion flows,\footnote{Quasi-spherical accretion flows may be relevant for relatively head-on stellar collisions where the neutron star becomes embedded fully within the disrupted stellar envelope or even for off-center tidal disruptions if the viscous evolution of the bound material ultimately causes the initially disk-like structure to ``puff up'' and become quasi-spherical \citep[e.g.,][]{Abramowicz1988,KingBegelman1999,McKinney2014,Dai2018}. On the other hand, initially quasi-spherical flows may in some cases become disk-like if net angular momentum is introduced as the compact object sweeps through material with a steep radial gradient of density \citep[e.g.,][]{Murguia-Berthier2017}.} where the geometry doesn't necessarily facilitate a low-density region through which accretion energy can escape unabsorbed, sufficiently high mass inflow rates render the Eddington limit irrelevant because photons can be trapped and advected inward with the accretion flow \citep[e.g.,][]{Rees1978,Begelman1979}. In this case, an accretion shock is expected to form near the neutron star surface within which the temperature and density are sufficiently high for neutrinos (produced through pair annihilation) to become the dominant cooling mechanism, carrying away the accretion energy without impeding the flow onto the neutron star \citep[e.g.,][]{Colgate1971}. In the spherical limit, previous studies \citep[e.g.,][]{Chevalier1993,Chevalier1996, Brown1995, FryerBenzHerant1996} have shown that for mass transfer rates above a critical value $\dot{M}_{\rm cr} \approx 10^{-4} M_{\odot}\,\rm{yr}^{-1}$, neutrino cooling allows for hypercritical accretion. For accretion flows with some rotational support,
$\dot{M}_{\rm cr}$ may be slightly larger than the spherical case \citep[e.g.,][]{Chevalier1993}, but hypercritical accretion is still expected to be possible \citep[e.g.,][]{Chevalier1996,ArmitageLivio2000,Brown2000,ZhangDai2009}. The physics of hypercritical accretion for neutrino-dominated accretion disks has also been explored at length in the context of black hole accretion, especially in the context of gamma-ray burst models \citep[e.g.,][]{Popham1999,Narayan2001,DiMatteo2002,Kohri2002,Lei2009}. In the black hole case, the critical mass transfer rate for neutrino cooling is estimated to be roughly $0.01\,M_{\odot}\,\rm{s}^{-1}$, much larger than in the neutron star case since in the former case accretion energy can disappear into the black hole.

With the above considerations in mind, we assume in what follows that the accretion flow is disk-like (motivated by the outcomes of our hydrodynamic simulations) and that hypercritical (e.g., super-Eddington) accretion onto the neutron star is possible at all times.

In order to take into account potential mass outflows expected in these hypercritical accretion disks \citep[e.g.,][]{NarayanYi1995,Blandford1999,Metzger2008}, we parameterize the mass inflow rate as a power-law in radius

\begin{equation}
    \label{eq:Macc}
    \dot{M} \approx \frac{M_{\rm{disk}}}{t_v} \Big( \frac{R_{\rm{acc}}}{R_{\rm{disk}}} \Big)^s 
\end{equation}
where $R_{\rm acc}$ is the accretion radius (i.e., the radius of the inner edge of the disk) and $t_v$ is the viscous accretion time. The exponent $s\in[0,1]$ parameterizes the uncertain amount of material transported from the outer edge of the disk (near the tidal disruption radius) to the accretion radius \citep{Blandford1999}. For example, for $s\approx 0.5$ \citep[e.g.,][]{Yuan2012} and for $R_{\rm acc}=10^6\,$cm and $R_{\rm{disk}}\approx 10^{11}\,$cm (Table \ref{table:sims}), $(R_{\rm acc}/R_{\rm{disk}})^{0.5}\approx \rm{a\,few}\times10^{-3}$. In this case, only a fraction of the disk mass is actually accreted and the remainder is ejected via a disk wind.

For the high magnetic field strengths expected in neutron stars, magnetic stresses can in principle dominate the flow in the accretion disk, especially near the neutron star surface where the field is strongest. The Alfv\'{e}n radius, $r_A$, is the characteristic distance from the neutron star at which the magnetic energy density is equal to the kinetic energy density of the material in the disk \citep[e.g.,][]{ShapiroTeukolsky1983}:

\begin{multline}
    r_A = \Bigg( \frac{\mu^4}{2 G M_{\rm NS} \dot{M}^2} \Bigg)^{1/7} \\ 
    \approx 1.9\times10^5\, \Big( \frac{B}{10^{11}\,\rm{G}} \Big)^{4/7} \Big( \frac{\dot{M}}{1\,M_{\odot}\,\rm{yr}^{-1}} \Big)^{-2/7}\,\rm{cm}
\end{multline}
where $\mu =B R_{\rm NS}^3$ is the assumed magnetic moment and where we have adopted $M_{\rm NS} = 1.2\,M_{\odot}$ and $R_{\rm NS}=10^6\,$cm. For high mass inflow rates representative of $s\lesssim 0.2$, the Alfv\'{e}n radius lies within the neutron star radius for field strengths expected for old neutron stars found in typical globular clusters \citep[e.g., $B\lesssim 10^{11}\,$G;][]{Ye2019}.
In this limit, the magnetic field is ``smothered'' by the large inflow of mass and the magnetic stresses do not play an important role in the accretion process. For lower mass transfer rates (or higher field strengths), the disk is truncated at $r_A$, and the accretion flow onto the neutron star surface is dominated by the magnetic field (i.e., the accretion flows onto the neutron star along the field lines in the magnetosphere). 

Following \citet{Metzger2008}, the time dependence of the accretion rate for thick disks parameterized by Equation~(\ref{eq:Macc}) can be expressed as
\begin{multline}
    \label{eq:Mdot}
    \dot{M}(t) \approx \Big( \frac{M_{d,i}}{t_{v,i}} \Big) \Big( \frac{R_{\rm acc}}{R_{d,i}} \Big)^s \times \\
    \Big[ 1 + 3(1-C)\Big(\frac{t}{t_{v,i}}\Big)\Big]^{-\frac{1+3(1+2s/3)(1-C)}{3(1-C)}}
\end{multline} where $C=2s/(2s+1)$ and $M_{d,i}$, $R_{d,i}$, and $t_{v,i}$ are the initial disk mass, disk radius, and viscous accretion time, respectively.\footnote{As in \citet{Metzger2008}, we have implicitly assumed that the disk wind outflow produces no net torque on the disk (i.e., the outflow carries away only the specific angular momentum of the mass lost). This assumption appears qualitatively consistent with global MHD disk simulations \citep[e.g.,][]{StonePringle2001}. If instead, the outflows do produce a significant net torque on the disk, $\dot{M}$ is expected to decrease much more rapidly \citep{Metzger2008}. This may steepen the decay of the lightcurve of an associated electromagnetic transient \citep[e.g.,][]{Metzger2021} and may potentially reduce the total mass accreted by the neutron star, potentially inhibiting the ability to produce a MSP. Of course in the case of $s\sim0$ where the disk outflow is negligible, the disk torques are also negligible and the conclusions here are unchanged.} $R_{\rm acc}$ is the accretion radius, which we define as the maximum of $R_{\rm NS}$ and $r_A$ at a given time. The total mass accreted by the neutron star after time $t$ can be computed by integrating Equation~(\ref{eq:Mdot}).

The time derivative of the total accreted angular momentum can be expressed as

\begin{multline}
    \label{eq:Jdot}
    \dot{J}_{\rm acc} \approx \frac{d}{dt}\Bigg( M_{\rm acc}\sqrt{GM_{\rm{NS}} R_{\rm acc}} \Bigg) \\
    \approx \dot{M}_{\rm acc} \Bigg( \sqrt{GM_{\rm{NS}} R_{\rm acc}} + \frac{1}{2}M_{\rm acc} \sqrt{\frac{G R_{\rm acc}}{M_{\rm NS}}} \Bigg)
\end{multline}
where $M_{\rm acc}$ is the total mass accreted after time $t$ and $M_{\rm NS}$ is the (evolving) neutron star mass. We have assumed that $\dot{M}_{\rm NS} = \dot{M}_{\rm acc}$ (since $\dot{M}_{\rm acc}$ already incorporates implicitly outflows associated with the $s$ parameter, this assumption is equivalent to stating simply that all material successfully transported to the neutron star surface is accreted) and that the accretion radius $R_{\rm acc}$ is roughly constant in time (especially appropriate for $r_A < r_{\rm NS}$ when most of the mass is accreted).
Integration of Equation~(\ref{eq:Jdot}) gives the total angular momentum supplied to the neutron star after time $t$.

In Figure~\ref{fig:disk}, we show the accretion rate, total accreted mass, and total accreted angular momentum for the neutron star versus time (with $t=0$ defined as the time of disk formation, i.e., roughly the end of the SPH simulations discussed in Section~\ref{sec:SPH}). As different colors, we show different assumed values of the uncertain $s$ parameter, ranging from $s=0$ (the case of highest mass inflow rate) to $s=1$ (a much lower mass inflow rate case, in which the majority of disk mass is blown away in a wind). In the left column, we show the evolution for an initial disk mass $M_{d,i}=0.4\,M_{\odot}$, initial disk radius $R_{d,i}=9\,R_{\odot}$, and viscous accretion time $t_{v,i}=5\,$d, representative of the tidal disruption of a $0.5\,M_{\odot}$ M-dwarf (e.g., simulation 3 in Table \ref{table:sims}). For the plots in the right column, we assume $M_{d,i}=1\,M_{\odot}$, $R_{d,i}=3\,R_{\odot}$, and $t_{v,i}=2\,$d, typical of the $1.2\,M_{\odot}$ BSS TDE case (e.g., simulation~6). In all cases, we assume a magnetic field strength of $10^{11}\,$G \citep[e.g.,][]{Ye2019}. Solid curves indicate evolution where the mass transfer rate is sufficiently high for the Alfv\'{e}n radius to lie within the neutron star radius. In this case the magnetic field is ``smothered'' and does not play an important role in the accretion flow. Dashed lines indicate the point in evolution where magnetic energy density dominates the accretion flow so $r_A>R_{\rm NS}$. In reality, the neutron star magnetic field may decrease (be ``buried'') through the accretion process \citep[e.g.,][]{Bhattacharya1991}, in which case the magnetic field may not influence the accretion flow until even later times still (at even lower accretion rates).

For a neutron star with moment of inertia $2M_{\rm NS} R_{\rm NS}^2/5$ and spin period $P_s$, the spin angular momentum is

\begin{equation}
    J_s = \frac{2}{5} M_{\rm NS} R_{\rm NS}^2 \Big( \frac{2\pi}{P_s} \Big).
\end{equation}
Assuming the majority of accretion occurs at the neutron star radius (appropriate for low $s$ cases as shown in Figure~\ref{fig:disk}) and $M_{\rm NS} \gg M_{\rm acc}$, the total accreted angular momentum can be expressed simply as $J_{\rm acc}\sim\,M_{\rm acc} \sqrt{G M_{\rm NS} R_{\rm NS}}$. In this case (assuming the initial spin period is roughly zero), the spin period attained through accretion is given roughly by

\begin{equation}
    \label{eq:spin}
    P_s \approx \frac{4 \pi}{5M_{\rm{acc}}} \Big( \frac{M_{\rm NS} R_{\rm NS}^3}{G} \Big)^{1/2}
    \approx 10 \Big( \frac{M_{\rm acc}}{0.02\,M_{\odot}} \Big)^{-1}\,\rm{ms}.
\end{equation}
Thus, the available disk mass is sufficient to spin up the neutron star to millisecond spin periods. The exact spin period attained depends on the efficiency of the accretion. We show as horizontal dotted lines in the bottom panels of Figure~\ref{fig:disk} the spin angular momentum values corresponding to a few characteristic neutron star spin periods (for $M_{\rm NS} = 1.2\,M_{\odot}$). As shown, in the high inflow case of $s\lesssim0.2$, a MSP can be produced for both disk masses assumed here.\footnote{In the most extreme case of $s=0$ for $M_{d,i}=1\,M_{\odot}$ (yellow curve in right hand panel of Figure \ref{fig:disk}), the total angular momentum accreted corresponds to a neutron star spin period of roughly $0.4\,$ms, near the break-up angular velocity of the neutron star (corresponding to the Keplerian velocity at the neutron star surface). In reality, once the break-up velocity is reached the neutron star is unlikely to be spun up further (although it may continue to accept mass).} 
For $s\gtrsim0.5$, the neutron star is unlikely to be spun up significantly, regardless of the disk mass.

Of specific relevance is the work by \citet{MacLeod2015} who evaluated the mass growth of a neutron star embedded within a common envelope \citep[taking into account the asymmetric structure of the structure of the envelope; e.g.,][]{MacLeod2015b}. Although a common envelope is not exactly identical to the accretion geometry expected for the tidal disruptions considered here, there are key qualitative similarities. This study found that, in general, a modest amount ($\lesssim0.1\,M_{\odot}$) of the envelope material is accreted by the neutron star. They pointed out that such mass growth is likely sufficient to spin up the neutron star (consistent with our result here), but is likely insufficient to lead to collapse to a black hole (a point we return to in Section \ref{sec:discussion}). 

\section{Discussion \& Conclusions}
\label{sec:discussion}

\begin{figure}
    \centering
    \includegraphics[width=0.95\columnwidth]{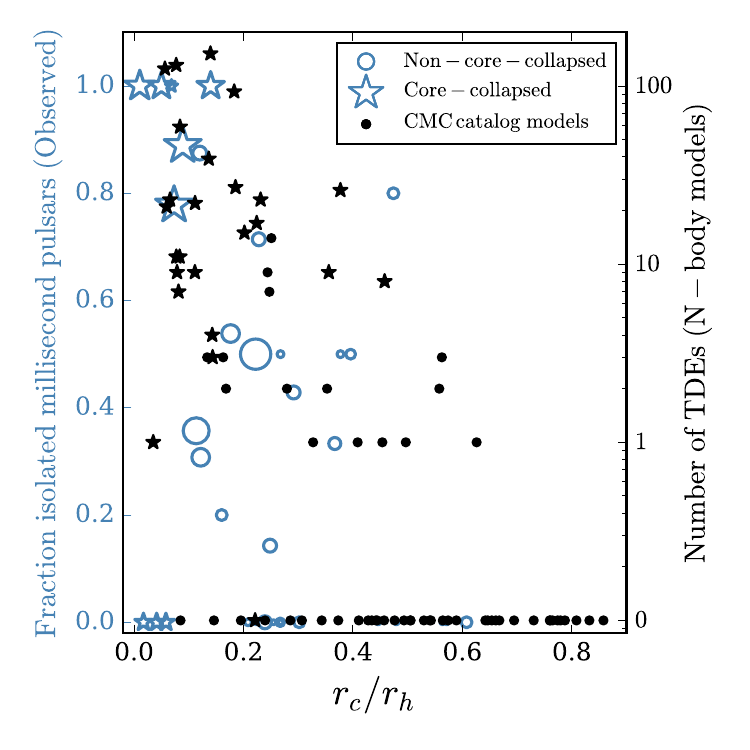}
    \caption{\footnotesize On the left-hand vertical axis (blue points) we show the fraction of isolated MSPs observed versus core radius over half-light radius for various Milky Way globular clusters. Open stars (circles) denote clusters that have (have not) undergone core collapse. The size of the points is scaled by the total number of MSPs (single or binary) observed in that cluster. On the right-hand vertical axis (black points), we show the total number of neutron star+star TDEs occurring in our various \texttt{CMC Catalog} cluster models (again black stars and circles denote core-collapsed and non-core-collapsed clusters, respectively). We argue the overabundance of isolated MSPs observed in the most centrally concentrated clusters can be explained in part by the increased rate of TDEs in these systems.}
    \label{fig:summary}
\end{figure}

Figure \ref{fig:summary} summarizes the key result of this study. On the left-hand vertical axis (blue color) we plot the fraction of observed isolated MSPs relative to the total number of observed MSPs (defined here as having spin periods less than $30\,$ms) versus core radius, $r_c$, over half-light radius, $r_h$, for all relevant Milky Way globular clusters. Blue stars indicate clusters that have undergone core collapse and blue circles indicate clusters that have not. The sizes of all blue points are scaled by the total number of MSPs observed in the cluster. On the right-hand vertical axis (black), we plot the total number of neutron star+main-sequence star TDEs versus $r_c/r_h$ for all relevant models from the \texttt{CMC Cluster Catalog} (again stars versus circles denote core-collapsed versus non-core-collapsed). As demonstrated in the figure, the most centrally dense clusters, especially those that have undergone core collapse, feature the highest fraction of isolated MSPs and the highest rate of TDEs. 

The choice of $30\,$ms to define a MSP is consistent with previous studies \citep[e.g.,][]{Lorimer2008,Ye2019}, but is admittedly somewhat arbitrary. Some core-collapsed clusters (e.g., M15 and NGC~6624) contain a handful of mildly recycled isolated pulsars with relatively large spin periods ($\gtrsim 100\,$ms) that in principle may also have formed through TDEs but with less mass accreted by the neutron star \citep[e.g.,][]{CamiloRasio2005}. Overall, the median spin period for all isolated pulsars in clusters is roughly $5.3\,$ms \citep{psr_catalog}. Less than $10\%$ of the observed isolated pulsars have periods in excess of $30\,$ms, thus this definition reasonably captures the bulk of the observed distribution. Additionally, we note that the spin period distribution for the observed \textit{binary} pulsars in clusters peaks at a slightly lower value than the isolated pulsars (the median spin period of the binary pulsars is roughly $3.7\,$ms) and a Kolmogorov–Smirnov test reveals these two distributions may in fact be distinct (KS statistic of roughly 0.3). Tentatively, this may hint at different formation channels for the binary versus isolated pulsars (e.g., classic binary mass transfer versus TDEs). 

As shown by Equation~(\ref{eq:spin}), the final spin period of the neutron star is determined by the mass accreted, which in turn is determined by the mass bound to the neutron star and the accretion efficiency of the disk (the $s$ parameter in Equation~\ref{eq:Macc}). 
For simplicity, we can assume that, as shown in our SPH models, roughly $90\%$ of the disrupted star becomes bound to the neutron star.
Under this assumption, from the distribution of stellar masses that undergo TDEs in our \texttt{CMC} models, we estimate that, for $s=0.2$ (assuming $R_{\rm disk} \approx 10^{11}\,$cm so that $[R_{\rm NS}/R_{\rm disk}]^{0.2}\approx0.1$), in {\em all\/} TDEs the neutron star would accrete sufficient mass to attain MSP spin periods ($P_s < 30\,$ms), corresponding to an average of roughly 10 isolated MSPs per cluster. For $s=0.4$ ($[R_{\rm NS}/R_{\rm disk}]^{0.4}\approx0.01$), roughly $56\%$ of TDEs (558 out of 988) would create MSPs, corresponding to roughly 5~isolated MSPs per cluster. For $s=0.5$ (roughly $0.3\%$ of the bound mass is accreted), only~41 (roughly 4\%) of all identified TDEs would lead to MSPs.

Current observations have revealed 98 isolated MSPs in the Milky Way globular clusters. Of these, 31 are observed in six core-collapsed clusters, 29 in the massive non-core-collapsed clusters Terzan~5 and 47~Tuc, and the remaining 38 are found in eleven lower-mass non-core-collapsed clusters. Because of various observational biases, this sample is likely to remain highly incomplete and the true number of isolated (as well as binary\footnote{Note that the observational biases are fewer and easier overcome for isolated pulsars than for binaries.}) MSPs could be much higher. In this case, the formation of roughly 5-10 isolated MSPs per cluster overall (including roughly 20-100 per typical core-collapsed cluster, roughly 20 per massive non-core-collapsed cluster like 47~Tuc or Terzan~5, and roughly 1 per typical low density cluster; see column 9 of Table \ref{table:CMC}) suggested by efficient disk accretion models is quite possibly consistent with observations.

As discussed in Section~\ref{sec:SPH}, regardless of the accretion and spin-up process, the neutron stars are expected to receive impulsive kicks of up to roughly $20\,$km/s from the asymmetric ejection of material stripped from the star during disruption. As a consequence of these kicks, we predict isolated MSPs formed through these TDEs should, on average, be found at large offsets from their host cluster's center. The average radial position (in units of their host cluster's core radius) of all the observed isolated MSPs in clusters with known radial positions is roughly 2.4 \citep{psr_catalog} -- these objects are clearly offset from their hosts' centers, which may be indicative of the velocity kicks proposed here.
For reference, this value for binary MSPs with known cluster positions is roughly 2, marginally lower than the isolated MSP value. 
We reserve for future study a detailed comparison of the observed offset distribution and the predicted offset distribution expected for the velocity kicks predicted by our models. 

Although we argue formation of isolated MSPs is a plausible outcome of neutron star TDEs, it is certainly not the only possibility. For inefficient accretion disks ($s\gtrsim0.2$) where only a small fraction of mass is accreted, the spin angular momentum of the neutron star will only increase slightly (Equation~\ref{eq:spin}). In this case, the TDEs would have a negligible effect upon the properties of the disrupting neutron stars even for the most massive disrupted stars. On the other hand, for highly efficient accretion disks ($s \sim 0$) that are sufficiently massive ($M_{\rm disk} \gtrsim 1.5 M_{\odot}$), the neutron star may accrete sufficient material to exceed the (uncertain) maximum allowable neutron star mass and, in this case, may collapse to form a {\em low-mass black hole\/}. The possibility of neutron stars being driven to collapse through accretion in a stellar envelope has been explored in the context of common envelope evolution of binaries \citep[e.g.,][]{Chevalier1993,BetheBrown1998,ArmitageLivio2000,Bethe2007,MacLeod2015}. In the specific case of a neutron star colliding with a massive main-sequence star,
where the final collision product qualitatively resembles a collapsar \citep{Woosley1993}, \citet{Hansen1998} showed that the collapse to a black hole may be accompanied by a (long) gamma-ray burst. For the case $s=0$ where the entire disk mass is accreted, we find that 39\% (16\%) of the TDEs in our \texttt{CMC} catalog models would lead to collapse to a low-mass black hole, assuming a maximum neutron star mass of $2\,M_{\odot}$ ($2.5\,M_{\odot}$). This translates to roughly $1-4$ low-mass black holes formed per typical core-collapsed cluster (in this case, the number of MSPs quoted previously would be reduced slightly since $16-39\%$ of the MSPs would instead become black holes). For $s\gtrsim0.2$, where less than 10\% of the disk mass is accreted, only one of the TDEs identified in our \texttt{CMC} models would lead to low-mass black hole formation. Therefore this outcome appears significant only if nearly the entire mass bound to the neutron star can be accreted.

A key process not considered here is the potential role of feedback energy in unbinding material initially bound to the neutron star (with binding energy $E_{\rm bind}\approx G M_{\rm NS}M_{\rm disk}/R_{\rm disk} \approx 10^{48} \,\rm{erg}$). In principle, material may be unbound before the roughly $10^{-2}M_{\odot}$ necessary to attain millisecond spin periods can be accreted by the neutron star, thus inhibiting the viability of these TDEs as a MSP formation mechanism. Feedback may arise through accretion energy \citep[e.g.,][]{ArmitageLivio2000,Papish2013} or nuclear energy generated through burning of hydrogen (and possibly heavier elements) near the neutron star surface \citep[e.g.,][]{Hansen1975,Taam1985,Bildsten1998}. Energy generated through accretion is expected to be of order $E_{\rm acc}\sim \eta M_{\rm acc} c^2$, where $\eta$ is the uncertain accretion efficiency. For a typical $\eta\approx 0.01$, $M_{\rm acc}\approx 10^{-4}\,M_{\odot}$ is sufficient to unbind the envelope, assuming the accretion energy can very efficiently  couple mechanically with the envelope. In reality, for disks similar to those considered here, a fraction of the accretion energy can likely be released relatively unabsorbed through the polar regions \citep[e.g., a jet-like geometry;][]{Livio1999}.

In the nuclear energy case, previous studies have demonstrated in the context of common envelope episodes, nuclear energy may be sufficient to eject remaining bound material \citep[e.g.,][]{Podsiadlowski2010,Ivanova2015,Grichener2018}. However, the efficiency of mechanical coupling is also key here; if the coupling is inefficient and most of the nuclear energy can be released as radiation \citep[e.g.,][]{Grichener2018,Soker2018} then ejection of the envelope may be difficult. We reserve for future study treatment of the possible role of feedback from both accretion and nuclear energy on the long-term outcome of these TDEs.


Given that the cross section for close encounters scales linearly with stellar radius in the parabolic regime, the occurrence of TDEs, with $r_p < r_T$, implies a comparable number of more distant encounters with $r_p$ in the range from $r_T$ to $\rm{a\,few}\times \it{r_T}$ that may form long-lived binaries through tidal capture \citep{Fabian1975}. In \citet{Ye2022}, we argued these tidal captures may eventually lead to the formation of ``redback'' MSPs \citep[e.g.,][]{Strader2019}, provided the companion star fills its Roche lobe and transfers mass onto (and spins up) the neutron star. This would imply an overabundance of redback MSPs in core-collapsed clusters, for the same reasons we argue tidal disruptions lead to an overabundance of isolated MSPs. There are 16~redbacks currently known in Milky Way globular clusters, four of which are in core-collapsed clusters.\footnote{Selection effects against finding redbacks are severe since they have large Doppler accelerations and long-duration and highly variable eclipses. These selection effects may become even more severe if the orbital periods are quite compact (i.e., $\mathcal{O}$[hour] as opposed to $\mathcal{O}$[day]). In this case, if short orbital period redbacks are produced by tidal capture, a large fraction of them may never be identified.} Given that only $20\%$ of Milky Way clusters are core-collapsed \citep{Harris1996}, 
this perhaps suggests a marginal overabundance. From a hydrodynamic perspective, it remains unclear whether the ultimate fate of tidal captures is indeed the formation of a detached binary \citep[e.g.,][]{CamiloRasio2005}
and, if so, whether the amount of mass transferred, is sufficient to spin up the neutron star to millisecond periods. Alternatively, depending on how quickly the tidally distorted star can radiate away the dissipated tidal energy, tidal captures may lead ultimately to mergers as we see clearly for closer encounters. In that case, the numbers of isolated MSPs predicted here may increase by a small factor. We reserve for future work more careful consideration of the distinction between tidal disruptions and captures and the implementation of a self-consistent treatment of the formation and fate of MSPs through tidal disruptions and captures within \texttt{CMC}.

As summarized in \citet{Kremer2021c}, the neutron star+main-sequence star TDEs considered here are just one of several possible processes expected in core-collapsed clusters that could in principle create rapidly spinning neutron stars. Accretion and spin-up may similarly occur for tidal disruptions of white dwarfs by neutron stars. As discussed in \citet{Metzger2012}, these TDEs may lead to orders-of-magnitude larger mass transfer rates. Concerning the event rates, on the one hand, the cross section for white dwarf tidal disruptions is a factor $\gtrsim100$ times smaller than for main-sequence stars (accounting for the relatively tiny radii but relatively high masses of white dwarfs compared to typical cluster main-sequence stars); on the other hand, white dwarfs are expected to be far more abundant in the inner regions of core-collapsed clusters \citep[e.g.,][]{Rui2021}. With this in mind, \citet{Kremer2021c} showed that the total rate of white dwarf + neutron star TDEs is roughly a factor of $10$ lower than the main-sequence star TDE rate. Alternatively, MSPs may be produced by mergers of pairs of white dwarfs \citep[e.g.,][]{Schwab2021,Kremer2021b}. Depending on various physical processes, white dwarf mergers may alternatively lead to Type~Ia supernovae \citep[e.g.,][]{Webbink1984}, slowly spinning pulsars possibly connected to the ``young pulsars'' observed in several Milky Way globular clusters \citep[e.g.,][]{Tauris2013}, or magnetars \citep[e.g.,][]{King2001}, which may also connect with fast radio bursts similar to FRB~20200120E \citep[e.g.,][]{Bhardwaj2021,Kremer2021b,Lu2022}. 
In an upcoming paper (Ye et al., in preparation) we will implement the formation of MSPs through neutron star + main-sequence star TDEs and other aforementioned mechanisms within \texttt{CMC}, enabling us to track self-consistently the formation and subsequent dynamical evolution of these objects.

\acknowledgments
We thank the anonymous referee for their careful review of the paper.
We also thank Tony Piro, Phil Hopkins, Nick Kaaz, and Ariadna Murguia-Berthier for helpful discussions. KK is supported by an NSF Astronomy and Astrophysics Postdoctoral Fellowship under award AST-2001751. FK acknowledges support from a CIERA Board of Visitors Graduate Fellowship. 
SMR is a CIFAR Fellow and is supported by the NSF Physics Frontiers Center awards 1430284 and 2020265. The National Radio Astronomy Observatory is a facility of the National Science Foundation operated under cooperative agreement by Associated Universities, Inc. This work was supported by NSF Grant AST-2108624 and NASA ATP Grant 80NSSC22K0722 at Northwestern University. 

\bibliographystyle{aasjournal}
\bibliography{mybib}

\end{document}